\documentclass{llncs}

\usepackage{graphicx}
\usepackage{booktabs}
\usepackage{url}
\usepackage[font=small,compatibility=false,tableposition=top]{caption}
\usepackage{floatrow}
\usepackage[pdfborder={0 0 0}]{hyperref}
\usepackage{subfig}
\usepackage{xspace}
\usepackage{amssymb}
\usepackage{amsmath}
\usepackage{gensymb}
\usepackage[mathcal]{euscript}
\usepackage[nolist,nohyperlinks]{acronym}
\usepackage[ruled,vlined,linesnumbered]{algorithm2e}
\usepackage{todonotes}
\usepackage{multirow}
\usepackage{hyphenat}
\usepackage[shortlabels]{enumitem}

\newcommand{\attar}{ATTar\xspace}

\ifdefined\algorithmautorefname
  \renewcommand{\algorithmautorefname}{Alg.\@}
\else
  
\fi

\newcommand{\ie}{{i.\,e.\@}\xspace}
\newcommand{\eg}{{e.\,g.\@}\xspace}
\newcommand{\etal}{{et al.\@}\xspace}

\acrodef{ilp}[ILP]{integer linear program}
\acrodef{fas}[FAS]{Feedback Arc Set}
\acrodef{fasp}[FASP]{Feedback Arc Set Problem}
\acrodef{mfas}[MFAS]{Minimum Feedback Arc Set}
\acrodef{scp}[SCP]{Sum Coloring Problem}
\acrodef{bscp}[BSCP]{Bandwidth Sum Coloring Problem}
\acrodef{dlp}[DLP]{Directed Layering Problem}
\acrodef{glp}[GLP]{Generalized Layering Problem}
\acrodef{lap}[LAP]{Linear Arrangement Problem}
\acrodef{olap}[OLAP]{Optimal Linear Arrangement Problem}
\acrodef{ldagp}[LDAGP]{Layered Directed Acyclic Graph with Ports}

\acrodef{ext}[EXT]{}
\acrodef{cglp}[CGLP]{Compact Generalized Layering Problem}
\acrodef{cgl}[CGL]{Compact Generalized Layering}
\acrodef{mml}[MML]{Min+Max Length}

\usepackage{mathtools}

\newcommand{\tsy}[0]{\textstyle}

\DeclareMathOperator{\rev}{rev}
\DeclareMathOperator{\len}{len}

\begin{document}
\mainmatter

\pagestyle{plain} %

\title{Compact Layered Drawings \\ of General Directed Graphs}

\author{
  Adalat Jabrayilov\inst{1}
  \and Sven Mallach\inst{3}
  \and Petra Mutzel\inst{1}
  \and\\ Ulf R{\"u}egg\inst{2}
  \and Reinhard von Hanxleden\inst{2}
}

\institute {
  Dept.\@ of Computer Science, Technische Universit{\"a}t Dortmund, \\
  Dortmund, Germany\\
  \email{\{adalat.jabrayilov,petra.mutzel\}@tu-dortmund.de}
  \and
  Dept.\@ of Computer Science, Kiel University, Kiel, Germany\\
  \email{\{uru,rvh\}@informatik.uni-kiel.de}
   \and
  Dept.\@ of Computer Science, Universit{\"a}t zu K{\"o}ln, K{\"o}ln, Germany\\
  \email{mallach@informatik.uni-koeln.de}
}

\maketitle

\setcounter{footnote}{0}

\begin{abstract}
We consider the problem of layering general directed graphs under height and possibly also
width constraints. Given a directed graph $G=(V,A)$ and a maximal height, we propose
a layering approach that minimizes a weighted sum of the number of reversed arcs, the arc
lengths, and the width of the drawing. We call this the \emph{\acf{cglp}}.
Here, the width of a drawing is defined as the maximum sum of the number of vertices
placed on a layer and the number of dummy vertices caused by arcs traversing the layer.
The CGLP is $\mathcal{NP}$-hard.
We present two MIP models for this problem.
The first one (EXT) is our extension of
a natural formulation for directed \emph{acyclic} graphs as suggested by Healy and Nikolov.
The second one (CGL) is a new formulation based on partial orderings.
Our computational experiments on two benchmark sets show that the CGL formulation can be
solved much faster than EXT using standard commercial MIP solvers. Moreover, we suggest
a variant of CGL, called MML, that can be seen as a heuristic approach.
In our experiments, MML clearly improves on CGL in terms of running time while it does
not considerably increase the average arc lengths and widths of the layouts
although it solves a slightly different problem where the dummy vertices are not taken into account.
\begin{keywords}
layer-based layout, layer assignment, mixed integer programming
\end{keywords}
\end{abstract}

\section{Introduction}
\label{sec:introduction}
\begin{figure}[tb]
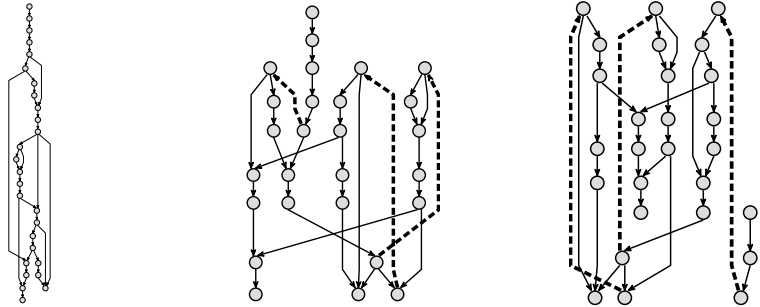

  \centering
  \begin{minipage}[b]{0.3\textwidth}
        \centering
    \subfloat[][Classic: 0 reversed arcs,\\44 dummy vertices]{
      \label{fig:mainexample_1}
      \begin{minipage}{\textwidth}
        \centering
        \includegraphics[scale=0.15]{{{data/layerings/drawings/g.27.4_d44_classic}}}
      \end{minipage}
    }
  \end{minipage}
\hfill
  \begin{minipage}[b]{0.3\textwidth}
        \centering
    \subfloat[][CGL: 3 reversed arcs,\\20 dummy vertices]{
      \label{fig:mainexample_2}
      \begin{minipage}{\textwidth}
        \centering
        \includegraphics[scale=0.35]{{{data/layerings/drawings/g.27.4_G}}}
      \end{minipage}
    }
      \end{minipage}
\hfill
  \begin{minipage}[b]{0.3\textwidth}
        \centering
    \subfloat[][MML: 3 reversed arcs,\\37 dummy vertices]{
      \label{fig:mainexample_3}
      \begin{minipage}{\textwidth}
        \centering
        \includegraphics[scale=0.35]{{{data/layerings/drawings/g.27.4_H}}}
      \end{minipage}
    }
  \end{minipage}
  \caption{
   (a) A graph
    drawn with traditional methods~\cite{EadesLS93,GansnerKNV93}
    where   %
    every arc has to point downwards, resulting in
    a poor aspect ratio -- here emphasized by scaling down
    the image to fit with the right ones.
    The methods CGL and MML, presented in this paper,
    are able to impose a bound on the height of the drawing,
    allowing some arcs to point upwards.
    The created drawings (b) and (c) are significantly more compact and
    improve readability. Reversed arcs are drawn bold and~dashed.
  }
  \label{fig:mainexample}
\end{figure}

A widely used hierarchical drawing style for directed graphs is the method proposed by Sugiyama \etal~\cite{SugiyamaTT81}
that involves the following steps in this order: (i) cycle removal, (ii) layering phase, (iii) crossing minimization, and (iv)
coordinate assignment and arc routing. One of the consequences of this workflow is that it is hard to control the aspect ratio
of the final layout. Phase (ii) requires an acyclic graph as input and the height of the produced layering inherently depends
on its longest path. So if phase (i) breaks cycles inappropriately or if an acyclic graph whose longest path is much larger than
its width is already given as initial input, it is impossible to construct a compact layering.
However, the readability and compactness of a drawing might be considerably improved if arcs to be reversed are chosen
carefully and in an integrated fashion, see \autoref{fig:mainexample} for an example.
If it is required that all arcs point downward, then the layering will have a poor aspect ratio as shown in
\autoref{fig:mainexample_1}. If we allow reversing some arcs so that they point upward, the aspect ratio can be
improved drastically (see \autoref{fig:mainexample_2} and \ref{fig:mainexample_3}).

In order to achieve this, R{\"u}egg \etal~\cite{RueeggESvH16} suggested to investigate the \emph{\acf{glp}}
that combines the first two interdependent phases of the Sugiyama approach. As the problem is $\mathcal{NP}$-hard, they
proposed an integer linear programming (ILP) formulation in which the weighted sum of the number of reversed arcs and the arcs
lengths is minimized, and where also the height
of the drawing can be restricted. Their approach improved the compactness of the derived layouts. However, it permits only
limited control over the width since it does not take the dummy vertices into account that are caused by arcs connecting
vertices on non-adjacent layers.
On the other hand, Healy and Nikolov~\cite{HealyN02b} have considered the layering problem with dummy vertices, but
only for acyclic directed graphs (DAGs) and for the case that both the desired maximum height and the width of the
drawing are given as inputs. %

\noindent\textit{Contributions.}
The purpose of this paper is to close this gap, i.e., to derive a model which is capable of computing a layering
of a general digraph by minimizing a weighted sum of the number of reversed arcs, the total arc length, and
the width $W$ of the resulting drawing taking the dummy vertices into account.
We call this the \emph{\acf{cglp}}.
The only input to our approach, besides the graph and the weights for the objective function, is
the desired maximum height $H$ of the drawing.
We will discuss how $H$ can be chosen such that the
existence of a feasible layering is always guaranteed
by our model --- which is in contrast to a setting where the user has to specify both $H$ and $W$.
Nevertheless, an upper bound on $W$ can still be specified. %

We present two mixed integer linear programming (MIP) formulations for the CGLP.
The first one (EXT) is a natural extension of the already mentioned model by
Healy and Nikolov for DAGs, and based on assignment variables.
The second one (CGL) is a completely new formulation based on partial orderings.
Our computational experiments on two benchmark sets show that the CGL formulation
can be solved much faster than EXT when using a standard commercial MIP solver
(Gurobi\footnote{\url{http://www.gurobi.com/}}) for both of the models.
Moreover, CGL is able to compute optimal solutions for each of the tested problem
instances in less than seven minutes of computation time on a standard PC while taking
only a few seconds for most of~them.

While in general we try to keep edges short,
we may want to emphasize reversed arcs,
for example in a flow diagram with cycles,
by drawing them rather long;
see also \autoref{fig:mainexample_3}.
For this reason, we propose a variant of CGL, called MML (\acl{mml}), for the problem of minimizing
the weighted sum of reversed arcs, positive forward arc lengths, negative backward arc lengths, and
the maximum number of real vertices on a layer. Within our experiments, MML is faster than both EXT
and CGL without considerably increasing the average arc lengths and widths of the layouts.

As mentioned by R{\"u}egg \etal, the aspect ratio of a final drawing (\ie in pixels)
does not only depend on the layering but also on the final phases of the Sugiyama
approach, \ie coordinate assignment and arc routing.
Even more, it can deviate significantly from the aspect ratio estimated
after the layering phase, as described in~\cite{RueeggESvH16}. Hence, in
practice, there is a high demand for methods that can be
quickly adjusted to the specific graph instance and use case.
Here, the models presented in this paper provide more control over
the produced layering compared to existing
approaches to the \acs{glp}.

\noindent\textit{Outline.}
The paper is organized as follows. First, we discuss related research in \autoref{sec:related_work}.
Definitions, preliminaries and motivations for our studies and models are given in \autoref{sec:definitions}.
\autoref{sec:original_algorithm} presents our newly developed MIP models which are finally evaluated
experimentally in \autoref{sec:evaluation}. We conclude with \autoref{sec:conclusion}.

\section{Related Work} \label{sec:related_work}

Over the years several approaches have been
proposed for the layering phase of the Sugiyama approach.
Eades and Sugiyama proposed
a method that is known
as longest path layering~\cite{EadesS90}.
This approach guarantees to produce a layering
with a minimum number of layers, \ie minimum height,
but the width can become arbitrarily large.
Gansner~\etal~\cite{GansnerKNV93}
use an \acs{ilp} formulation
to create layerings with minimum %
total arc length that can be solved efficiently using a
network simplex algorithm.
The Coffman-Graham algorithm~\cite{CoffmanG72}
delivers approximate solutions to
the precedence-constrained multi-processor
scheduling problem that can be used
to calculate a layering with a maximum number
of real vertices per layer.
However, dummy vertices are not taken into
account but can have a significant impact
on the actual width of the drawing.
Still, Nachmanson~\etal~\cite{NachmansonRL08}
use the Coffman-Graham algorithm as part of an
iterative heuristic procedure to produce
drawings with a certain aspect ratio.

The layering problem with restricted width
and consideration of dummy vertices
has been studied in the literature as well.
Healy and Nikolov found that minimizing the
number of dummy vertices during layering
inherently produces compact drawings~\cite{HealyN02a}.
Following this observation,
they target dummy vertex minimization with
a branch and cut algorithm that is
able to incorporate %
bounds on both width and height~\cite{HealyN02b}.
Nikolov~\etal
discuss heuristics to find layerings
with small width when considering
dummy vertices~\cite{NikolovTB05}, however,
no explicit bound on the width and
height can be used.

Since all of the above methods
rely on the input graph to be acyclic,
the minimum height of their layerings
directly relates to the longest
path of the graph.
R\"{u}egg~\etal~\cite{RueeggESvH16}
integrate the first two phases
of the Sugiyama approach to allow
arbitrary graphs as input
by minimizing a weighted sum
of the number of reversed arcs
and the number of dummy vertices.
They showed that this
overcomes the previously mentioned
problem regarding a graph's longest path
and also allows more compact drawings
in general.
Still, they did not consider
hard bounds on the width and height
of the drawing and did not consider
the contribution of dummy vertices to
the width.

\section{Preliminaries} \label{sec:definitions}

As we are concerned with the layering phase of the layout method suggested
by Sugiyama~\etal~\cite{SugiyamaTT81}, we briefly recall the definition of layerings of
directed graphs at the beginning of this section. Afterwards, we state our new layering
problem definition and discuss how to guarantee the existence of feasible drawings under
this model.

\paragraph{Layerings of directed graphs.}
A \emph{layering} of a directed graph $G=(V,A)$ with vertex set $V$ and arc set $A$
is a function $\ell: V \rightarrow \mathbb{N}^+$ assigning a layer $\ell(v)$ to each
vertex $v\in V$. In our context, for a feasible layering, it is necessary that no
two adjacent vertices are placed on the same layer, i.e., $\ell(u) \neq \ell(v)$ for
all $(u,v) \in A$. The layering determines the height, the width, the arc lengths, and
the number of reversed arcs, which will be defined in the following.
The \emph{height} $H$ of a layering~$\ell$ is the maximum layer used by $\ell$.
The \emph{width} $W_k(\ell)$ of a layer $k$ in a layering~$\ell$ is the sum of
the number of vertices assigned to layer $k$ and the number of dummy vertices caused by
arcs traversing the layer $k$. More formally, let ${V_k = \{ v\in V \mid \ell(v) = k \}}$
and $D_k = \{ (u,v) \in A \mid \ell(u) < k \mbox { and } \ell(v) > k \mbox { or } \ell(v) < k \mbox { and } \ell(u) > k \}$.
Then $W_k(\ell) = | V_k | + | D_k |$ and the width of $\ell$ is defined as the maximum width
of all layers in $\ell$, i.e., $W = \max_{1 \le k \le H} W_k(\ell)$.
The total arc length $\len(\ell)$ is the sum of the arc lengths
$|\ell(u) - \ell(v)|$ of all arcs $(u,v) \in A$ in the layering $\ell$.
An arc $(u,v)\in A$ in a layering $\ell$ is called a \emph{reverse arc} if $\ell(u)>\ell(v)$
and the total number of reversed arcs in $\ell$ is denoted by $\rev(\ell)$.
The \emph{estimated aspect ratio} of a layering with width $W$ and height $H$ is defined as $W/H$.
In contrast, the \emph{aspect ratio}, considers the width and height of a final layout after all of the Sugiyama phases.
So far, we assumed that real as well as
dummy vertices have unit width. Nevertheless, each of our MIP models presented
in~\autoref{sec:original_algorithm} can easily be extended to deal with varying
vertex widths.

Asking the user to specify bounds on both the height $H$ and the width $W$ of a
hierarchical drawing a priori can easily lead to infeasible problem settings or
require several iterations to fit the parameters to the
graph structure. We circumvent these issues in our subsequently defined
variation of the problem by requiring only $H$ as an input parameter while
making $W$ a subject of optimization. We also discuss how $H$ can be chosen safely.

\paragraph{The Compact Generalized Layering Problem (CGLP).}

As an extension to the Generalized Layering Problem (GLP) described in~\cite{RueeggESvH16}, we
define the Compact Generalized Layering Problem as follows:
Given a (not necessarily acyclic) directed graph $G=(V,A)$ and a maximum layering
height $H$, compute a layering~$\ell$ such that the end-vertices of each arc are
assigned to different layers and the following objective function is minimized:
the weighted sum of the number of reversed arcs $\rev(\ell)$, the total arc
length $\len(\ell)$, and the width $W(\ell)$.

\paragraph{Lower Bounds on the height $H$ of feasible layerings.}
Since, in our generalized setting, the direction of the arcs can be arbitrary, we
can think of undirected graphs to determine a lower bound on $H$. To assign layers
to vertices such that no two adjacent vertices are on the same layer is an equivalent
problem as to assign colors to vertices such that no two adjacent vertices have the
same color. Hence, the minimum number of layers, i.e., height, necessary for a feasible
layering of an undirected graph $G=(V,E)$ is equal to its chromatic number $\chi(G)$.

Since it is $\mathcal{NP}$-hard to compute $\chi(G)$, we suggest to approximate
it. A valid upper bound for $\chi(G)$ that can be computed in linear
time is the maximum vertex degree $\max_{v \in V} \deg(v)$ of $G$ plus one.
This bound is tight if $G$ is the complete graph or an odd cycle, otherwise
$\chi(G) \le \max_{v \in V} \deg(v)$ for any connected graph~$G$.
A better approximation can be achieved by using the largest
eigenvalue $\lambda^*$ of the adjacency matrix of $G$.
Wilf showed that $\chi(G) \le 1 + \lambda^*$~\cite{Wilf67} and together
with the Perron-Frobenius Theorem we have that
$2\frac{|E|}{|V|} \le \lambda^* \le \max_{v \in V} \deg(v)$.
Summing up, we have $H \ge \chi(G) \le \lambda^* + 1 \le \max_{v \in V} \deg(v) + 1$.

\section{Description of the MIP Models} \label{sec:original_algorithm}

\subsection{A Generalization of the Model by Healy and Nikolov (EXT)} \label{subsec:EXTmodel}

As a reference, we consider a natural extension of the model
by Healy and Nikolov~\cite{HealyN02b} that minimizes the total arc length
of layered drawings of \emph{acyclic} digraphs while restricting
the height as well as the width and taking dummy vertices into account.
In their model, the height $H$ and the width $W$ are fixed input parameters.
In our extension for general digraphs $G=(V,A)$, we only require $H$ as an
input, and incorporate the width of the drawing into the optimization process.
The model describes an assignment problem (AP) with variables $x_{v,k}$
to decide whether vertex $v \in V$
is placed on layer $1 \le k \le H$ ($x_{v,k} = 1$) or not ($x_{v,k} = 0$).
If an arc $(u,v) \in A$ is reversed, then this is expressed by a variable
$r_{u,v} = 1$, otherwise $r_{u,v} = 0$. The variables $z_{uv,k}$ model whether
arc $(u,v)$ causes a dummy vertex on layer $2 \le k \le H-1$, which is important to formulate
a proper width constraint. To save extra arc length variables, we exploit the
fact that the length of an arc is exactly the number of dummy
vertices it causes plus one. The full model~is:

\begin{small}
\begin{align}
    \min\;     &  \big(\omega_{rev} \tsy\sum\limits_{(u,v) \in A} r_{u,v}\big) + \big(\omega_{len} \tsy\sum\limits_{(u,v) \in A} \tsy\sum\limits_{k=2}^{H-1} z_{uv,k}\big) + \omega_{wid}\; W \span \span \span \span  & & \nonumber \\
s.t.\;\;   & \tsy\sum\limits_{k=1}^{H} x_{v,k}           &=\;   & 1            && \mbox{for all } v \in V \label{con:assignD} \\
           & x_{u,k} + x_{v,k}                             &\le\; & 1            && \mbox{for all } (u,v) \in A, 1 \le k \le H \label{con:noshareD} \\
	   & x_{u,k} - \tsy\sum\limits_{l=k}^{H} x_{v,l}&\le\; & r_{uv}       && \mbox{for all } (u,v) \in A, 1 \le k \le H \label{con:changeD} \\
	   & \tsy\sum\limits_{v \in V} x_{v,k}                                                      &\le\; & W            && \mbox{for all } k \in \{1, H\} \label{con:widthD1} \\
	   & \tsy\sum\limits_{v \in V} x_{v,k} + \tsy\sum\limits_{\mathclap{(u,v) \in A}} z_{uv,k}  &\le\; & W            && \mbox{for all } 2 \le k \le H-1 \label{con:widthD} \\
           & \tsy\sum\limits_{l < k} x_{v,l} - \tsy\sum\limits_{l \le k} x_{u,l}         &\le\;   & z_{uv,k}     && \mbox{for all }  (u,v) \in A, 2 \le k \le H-1 \label{con:dummy1} \\
           & \tsy\sum\limits_{l < k} x_{u,l} - \tsy\sum\limits_{l \le k} x_{v,l}         &\le\;   & z_{uv,k}     && \mbox{for all }  (u,v) \in A, 2 \le k \le H-1 \label{con:dummy2} \\
           & x_{v,k}                                       &\in\; & \{0,1\}    && \mbox{for all } v \in V, 1 \le k \le H \nonumber\\
	   & r_{u,v}                                       &\in\; & [0,1]      && \mbox{for all } (u,v) \in A \nonumber\\
	   & z_{uv,k}                                      &\in\; & [0,1]      && \mbox{for all } (u,v) \in A, 2 \le k \le H-1  \nonumber\\
	   & W                                             &\in\; & \mathbb{R}_{\ge 0}   && \nonumber
\end{align}
\end{small}
\noindent The objective function minimizes the weighted sum of the
number of reversed arcs, the total arc length, and the width $W$ of the
drawing. Equations~(\ref{con:assignD}) ensure that exactly one
layer is assigned to each $v \in V$. Inequalities~(\ref{con:noshareD})
enforce adjacent vertices to be placed on different layers. If an
arc $(u,v)$ is reversed due to the positions of its end vertices, then
inequality~(\ref{con:changeD}) makes sure
that $r_{u,v}$ is equal to $1$ (otherwise, it will be $0$ due to
the objective function). The total number of vertices and dummy
vertices assigned to one layer must never exceed $W$ which is
ensured by (\ref{con:widthD1}) and (\ref{con:widthD}).
Finally, inequalities~(\ref{con:dummy1})
and (\ref{con:dummy2}) enforce a variable $z_{uv,k}$ to be $1$ if $\ell(u) < k$
and $\ell(v) > k$ or vice versa. The integrality of all
continuous variables is implied by the integrality of the $x$-variables
due to the constraints and the objective~function.
Model EXT has $\mathcal{O}(|V| \cdot H + |A| \cdot H)$ variables and
$\mathcal{O}(|V| + |A| \cdot H)$ constraints.

\subsection{Our New Ordering-Based MIP Model (CGL)}
\label{subsec:CGLmodel}
The CGL model is based on the observation that the layering
problem is a partial ordering problem (POP) in the sense that a vertex $u$
is smaller than $v$ (i.e., $u<v$) in the partial order if $\ell(u) < \ell(v)$.
Following this idea, we introduce, for each $v \in V$ and for each $1 \le k \le H$, the variables
$y_{v,k}$ that are equal to $1$ if and only if $\ell(v) < k$.
Conceptually, we also have the reverse variables $y_{k,v}$ (equal to $1$ if
and only if $k < \ell(v)$), but we will see soon that these can be discarded.
However, with the reverse variables at hand, it is easy to see that
$\ell(v) = k$ if and only if $y_{k,v} = y_{v,k} = 0$.
In addition, the new model also comprises the variables $r_{u,v}$ and
$z_{uv,k}$ as already introduced in \autoref{subsec:EXTmodel}. The interplay
between the $y$- and the $r$-variables as described in the following will
lead to the desired partial ordering of $V$. The full model~is:

\begin{small}
\begin{align}
    \min\;     &  \big(\omega_{rev} \tsy\sum\limits_{(u,v) \in A} r_{u,v}\big) + \big(\omega_{len} \tsy\sum\limits_{(u,v) \in A} \tsy\sum\limits_{k=2}^{H-1} z_{uv,k}\big) + \omega_{wid}\; W \span \span \span \span  & & \nonumber \\
    s.t.\;\;     &  y_{v,1}                          &=\;   &  0         && \mbox{for all } v \in V   \label{cgl:none_below1}      \\
                 &  y_{H, v}                         &=\;   &  0         && \mbox{for all } v \in V   \label{cgl:none_aboveH}      \\
		 &  y_{k,v} + y_{v,k+1}                   &=\;   &  1         && \mbox{for all } v \in V , 1 \le k \le H-1 \label{cgl:xor}  \\
		 &  y_{{k+1}, v} - y_{k,v}                &\le\; &  0         && \mbox{for all } v \in V , 1 \le k \le H-2 \label{cgl:trans} \\
		 &{-y_{u,k}} - y_{k,v} - r_{u,v}          &\le\; & {-1}       && \mbox{for all } (u,v) \in A, 1 \le k \le H  \label{cgl:setR1}\\
		 &{-y_{k,u}} - y_{v,k} + r_{u,v}          &\le\; &  0         && \mbox{for all } (u,v) \in A, 1 \le k \le H  \label{cgl:setR4}\\
                 & y_{k,u} + y_{v,k} -      z_{uv,k}      &\le\; &  1         && \mbox{for all } (u,v) \in A, 2 \le k \le H-1 \label{cgl:dummy1}\\
                 & y_{k,v} + y_{u,k} -      z_{uv,k}      &\le\; &  1         && \mbox{for all } (u,v) \in A, 2 \le k \le H-1 \label{cgl:dummy2}\\
    & \tsy\sum\limits_{\mathclap{u \in V}} ( 1 - y_{u,k} - y_{k,u} )  &\le\; & W            && \mbox{for all } k \in \{1,H\} \label{cgl:width1} \\
    & \tsy\sum\limits_{\mathclap{u \in V}} ( 1 - y_{u,k} - y_{k,u} ) + \tsy\sum\limits_{\mathclap{(u,v) \in A}} z_{uv,k} &\le\; & W            && \mbox{for all } 2 \le k \le H-1 \label{cgl:width} \\
		 & y_{v,k},\; y_{k,v}                     &\in\; & \{0,1\}    && \mbox{for all } v \in V, 1 \le k \le H \nonumber\\
                 & r_{u,v}                                &\in\; & [0,1]      && \mbox{for all } (u,v) \in A \nonumber\\
                 & z_{uv,k}                               &\in\; & [0,1]      && \mbox{for all } (u,v) \in A, 2 \le k \le H-1  \nonumber\\
                 & W                                      &\in\; & \mathbb{R}_{\ge 0}   && \nonumber
\end{align}
\end{small}

\noindent The switch from an AP to a POP requires a more
involved approach to yield consistency of the model.
The first four constraints enforce the graph to be embedded into the layers $1, \dots, H$.
Equations~(\ref{cgl:none_below1}) and~(\ref{cgl:none_aboveH})
make sure no vertex is assigned a layer smaller than one or larger than $H$.
For each layer $1 \le k \le H$, each vertex $v$ is either assigned a layer larger
than $k$ (in which case $y_{k,v} = 1$) or not (in which case $y_{v,k+1} = 1$)
as is enforced by~(\ref{cgl:xor}). These equations can be used to eliminate one
half of the $y$-variables (and then be eliminated themselves) as mentioned before.
If $\ell(v) > k + 1$, then this implies $\ell(v) > k$ as well and this is
expressed in the transitivity inequalities~(\ref{cgl:trans}). It remains to
show that arc directions and layer assignments will be consistent and no
two adjacent vertices can be on the same layer. This is achieved by
inequalities (\ref{cgl:setR1}) and (\ref{cgl:setR4}).
Suppose that
$r_{u,v} = 0$, i.e., the arc $(u,v) \in A$ shall be a forward arc.
Then the inequalities (\ref{cgl:setR1}) enforce that, for each layer
$1 \le k \le H$, either $\ell(u) < k$ or $\ell(v) > k$ (or both).
In this case, the inequalities~(\ref{cgl:setR4}) %
are inactive,
but they take the equivalent role in the reversed-arc case where $r_{u,v} = 1$
and then inequalities~(\ref{cgl:setR1}) %
are inactive.

As already discussed for the previous model, a dummy vertex on layer $k$ is
caused by arc $(u,v) \in A$ if either $\ell(u) > k$ and $\ell(v) < k$
($y_{k,u} + y_{v,k} - 1 = 1$), or vice versa ($y_{k,v} + y_{u,k} - 1 = 1$).
In the first case, inequality~(\ref{cgl:dummy1}) will force $z_{uv,k}$ to be $1$,
in the second case, inequality~(\ref{cgl:dummy2}) will do so. In any other case,
the variable will be zero due to the objective function.
Finally, inequalities~(\ref{cgl:width1}) and (\ref{cgl:width}) count the vertices
and dummy vertices placed on each layer $k$ and make sure that $W$ is a
proper upper bound on the width of the layering.
The CGL formulation has $\mathcal{O}(|V| \cdot H + |A| \cdot H)$ variables and
constraints.

\subsection{A Min+Max Length Variant Without Dummy Vertices (MML)}
\label{subsec:MMLmodel}

We shortly describe a variant of CGL, called MML, that can produce
appealing results usually faster when dummy vertices need not be taken
into account in terms of the~width. As opposed to $W$, let $W_r$ be the
width of a layering where only the real vertices are~counted.

The idea is to remove the dummy vertex variables $z_{uv,k}$
together with the constraints (\ref{cgl:dummy1}), (\ref{cgl:dummy2}) and
to replace inequalities (\ref{cgl:width1}) and (\ref{cgl:width}) simply by:
\begin{small}
\begin{align}
    & \tsy\sum\limits_{\mathclap{u \in V}} ( 1 - y_{u,k} - y_{k,u})  &\le\; & W_r  && \mbox{for all } 1 \le k \le H \nonumber
\end{align}
\end{small}

We now need to count arc lengths in an ordinary fashion.
The usual way to do this is to introduce
length variables $l_{u,v} \in \mathbb{R}$ and the following two inequalities
per arc in order to capture the absolute
length depending on the arc direction.
\begin{small}
\begin{align}
    & \tsy\sum\limits_{k=1}^{H} (y_{k,v} - y_{k,u}) &\le\; &  l_{u,v}   && \mbox{for all } (u,v) \in A \label{MML:length1}\\
    & \tsy\sum\limits_{k=1}^{H} (y_{k,u} - y_{k,v}) &\le\; &  l_{u,v}   && \mbox{for all } (u,v) \in A \label{MML:length2}
\end{align}
\end{small}

However, for certain use cases,
\eg, when feedback should be emphasized,
it can be desirable
to draw forward arcs as short as possible
and maximize the length of the reversed arcs.
Therefore, we propose not to introduce the $l$-variables but to directly
incorporate the terms used on the left hand side of inequalities~(\ref{MML:length1}) into
the objective function,
which results in the desired minimization of the
backward arcs' negative lengths.
MML's objective function is:
\begin{small}
$$\min\; \big(\omega_{rev} \tsy\sum\limits_{(u,v) \in A} r_{u,v}\big) + \big(\omega_{len} \tsy\sum\limits_{(u,v) \in A} \tsy\sum\limits_{k=1}^{H} (y_{k,v} - y_{k,u})\big) + \omega_{wid}\; W_r$$
\end{small}

This model has only $\mathcal{O}(|V| \cdot H + |A|)$
variables and $\mathcal{O}(|V| \cdot H + |A| \cdot H)$ constraints.

\section{Evaluation}
\label{sec:evaluation}

\paragraph{Setup.}
The experiments were performed single-threadedly on an Intel
Core i7-4790, 3.6~GHz, with 32 GB of memory and running Ubuntu
Linux 14.04. For solving the MIPs, we used Gurobi 6.5.
In our implementation, the CGL model has been reduced in terms
of its variables as is indicated in~\autoref{subsec:CGLmodel}.
Further, in all the models we enforce at least one vertex to
be placed on layer $k=1$ to eliminate some symmetries.
The parameters were set to ${H=\lceil 1.6*\sqrt{|V|} \rceil}$, $w_{rev} = |E| \cdot H$, $w_{len} = 1$, and
$w_{wid} = 1$.  This choice of $H$ delivered feasible
problems for all of our instances and emphasizes our target
to have a good aspect ratio and a drawing that adheres to
standard forms such as flat screens following the golden ratio.
Due to our choice of $w_{rev}$, arcs are reversed only if
this is unavoidable due to the specified height or because they are part of a cycle.
We are interested in answering the following questions:
\begin{enumerate}
\item[(H1)] Does the POP-oriented CGL model dominate the AP-based EXT model in terms of running times?
\item[(H2)] Is MML a good alternative concerning the running times and the metrics (arc length, $W$, and estimated aspect ratio) of the generated layerings?
\item[(H3)] How do CGL and MML influence the aspect ratio of the \emph{final} drawings?
\end{enumerate}

We used two benchmark sets\footnote{Both benchmark sets are available on \url{https://ls11-www.cs.tu-dortmund.de/mutzel/gdbenchmarks}}.
The first set \emph{\attar}, the same as used by R\"{u}egg~\etal~\cite{RueeggESvH16}, is an extraction of $146$ acyclic AT\&T graphs
with at least $20$ vertices having aspect ratio smaller than
$0.5$ when drawn with the classic Sugiyama approach. %
These graphs have between $20$ and $99$ vertices and between $20$ and $168$ arcs.
Their arc to vertex density is about $1.5$ on average but varies significantly.
Especially, the density of some graphs with about $60$ vertices is up to $4.7$
which is why the results displayed in our figures and boxplots stand out
for these instances.

The second benchmark set \emph{Random} consists of $340$ randomly generated,
not necessarily acyclic graphs with $17$ to $100$ vertices,
$30$ to $158$ arcs, and $1.5$ arcs per vertex.
We used these graphs in order to analyze our approach also for
cyclic sparse digraphs.
First, a number of vertices was created. Afterwards,
for each vertex, a random number of outgoing arcs
(with arbitrary target) is created such that
the overall number of arcs is $1.5$ times the number of vertices.

\begin{figure}[tb]
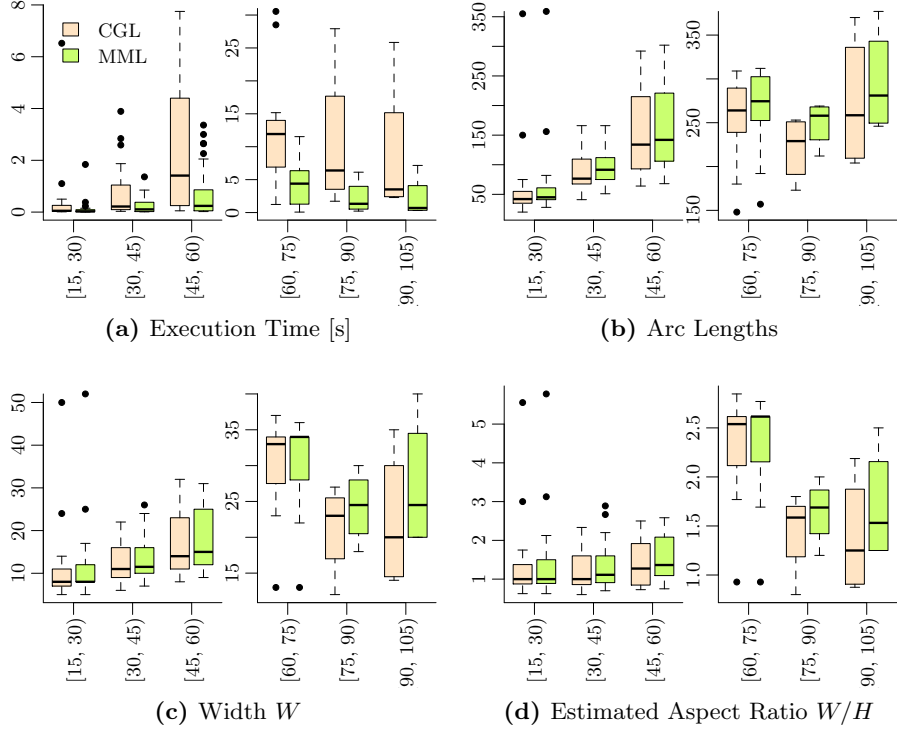

  \centering
  \subfloat[Execution Time {[s]}]{
    \label{fig:results_summary_att_exec}
    \resizebox{.48\textwidth}{!}{%
      \input{plots/boxplot_binned_time.sec._att.tikz}
    }
  }\hfill
  \subfloat[Arc Lengths]{
    \label{fig:results_summary_att_len}
    \resizebox{.48\textwidth}{!}{%
      \input{plots/boxplot_binned_len_att.tikz}
    }
  }\\
  \subfloat[Width $W$]{
    \label{fig:results_summary_att_w}
    \resizebox{.48\textwidth}{!}{%
      \input{plots/boxplot_binned_W_att.tikz}
    }
  }\hfill
  \subfloat[Estimated Aspect Ratio $W/H$]{
    \label{fig:results_summary_att_ar}
    \resizebox{.48\textwidth}{!}{%
      \input{plots/boxplot_binned_ar_att.tikz}
    }
  }
  \caption{
    Summary of the results of the \attar graphs.
    For each of the four metrics
    the graphs were binned based on their vertex counts (x axis)
    and the y axis represents the metric's value.
    For each bin the left box represents CGL's result
    and the right box MML's result.
  }
  \label{fig:results_summary_att}
\end{figure}

\begin{figure}[tb]
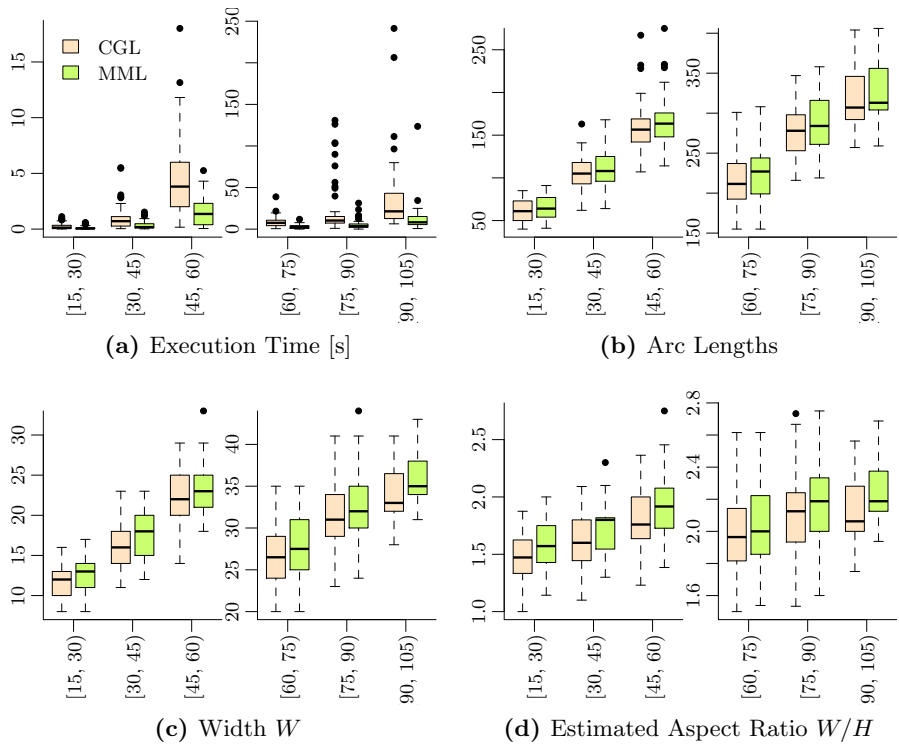

  \centering
  \subfloat[Execution Time {[s]}]{
    \label{fig:results_summary_random_exec}
    \resizebox{.48\textwidth}{!}{%
      \input{plots/boxplot_binned_time.sec._rand.tikz}
    }
  }\hfill
  \subfloat[Arc Lengths]{
    \label{fig:results_summary_random_len}
    \resizebox{.48\textwidth}{!}{%
      \input{plots/boxplot_binned_len_rand.tikz}
    }
  }\\
  \subfloat[Width $W$]{
    \label{fig:results_summary_random_w}
    \resizebox{.48\textwidth}{!}{%
      \input{plots/boxplot_binned_W_rand.tikz}
    }
  }\hfill
  \subfloat[Estimated Aspect Ratio $W/H$]{
    \label{fig:results_summary_random_ar}
    \resizebox{.48\textwidth}{!}{%
      \input{plots/boxplot_binned_ar_rand.tikz}
    }
  }
  \caption{
    Summary of the results of the random graphs.
    For each of the four metrics, the graphs
    were binned based on their vertex counts (x axis)
    and the y axis represents the metric's value.
    For each bin, the left box represents CGL's result
    and the right box MML's result.
    To improve presentation, we removed one outlier in (a)
    with $92$ vertices that took CGL $388s$ of computation time.
  }
  \label{fig:results_summary_random}
\end{figure}

\paragraph{(H1): Comparison of CGL and EXT.}
First, we look at the \attar instances and model EXT. While small instances with
up to $25$ vertices can be solved within at most seven seconds, there is already
one instance with $29$ vertices that cannot be solved within the time limit of $10$
minutes of CPU and system time. In total, eleven of $66$ instances with ${25 < |V| < 50}$
time out, while for the others the running times highly deviate within the full spectrum
between a second and about nine minutes. However, none of the $33$ instances with more
than $50$ vertices can be solved within the time limit.
The picture for the random instances is similar. The first instance remaining unsolved
within the time limit has $37$ vertices and those with $50$ or more vertices can
be solved only sporadically ($33$~of~$201$).

With the CGL model, however, we were able to solve all the instances (\attar and Random)
to optimality. The running times are shown
in \autoref{fig:results_summary_att_exec}~and~\ref{fig:results_summary_random_exec}.
All but ten of the \attar instances were solved in less than $10$ seconds of CPU and system time,
and the highest running time observed was $30$ seconds.
Concerning the random instances, we observed that $314$ of the $340$ instances were
solved within $30$ seconds. However, for $|V|\ge 80$, a higher dispersion of
running times could be observed.
The largest observed running time was $388$ seconds for an instance with 92 vertices.
Since both models solve the same problem to optimality, we can conclude
that the EXT model is clearly dominated by the CGL model when a state-of-the-art commercial
MIP solver is used.

\paragraph{(H2): Alternative MML.}

In \autoref{subsec:MMLmodel} we introduced
the model MML that, as opposed to the models EXT and CGL,
maximizes the length of reversed arcs
and does not regard the contribution
of dummy vertices to a layer's width.
Our hope was that MML is much faster than CGL and EXT without sacrificing the quality of the generated layouts too much.

In \autoref{fig:results_summary_att_exec}~and~\ref{fig:results_summary_random_exec},
one can see that the MML model
could almost always (except for $14$ instances in total) be solved much faster
than the CGL model and hence also the EXT model.
Especially, with MML all but ten instances of the \attar benchmark set
were solved within three seconds.
The boxplots also show that the running times for solving the MML model
are more robust for both benchmark sets.

As can be seen in \autoref{fig:results_summary_att} and \ref{fig:results_summary_random},
the average arc lengths and widths of layerings created by MML
increase only moderately when compared to CGL.
On average, the increase in the total arc lengths is only about $6\%$ and the increase in
the width is about $7\%$ for each of the benchmark sets.
The displayed widths of MML also include dummy vertices.
Since MML is significantly faster than CGL, we can conclude that it is
a good alternative to CGL
whenever lower running times are required
and long reversed arcs are either desired or negligible.

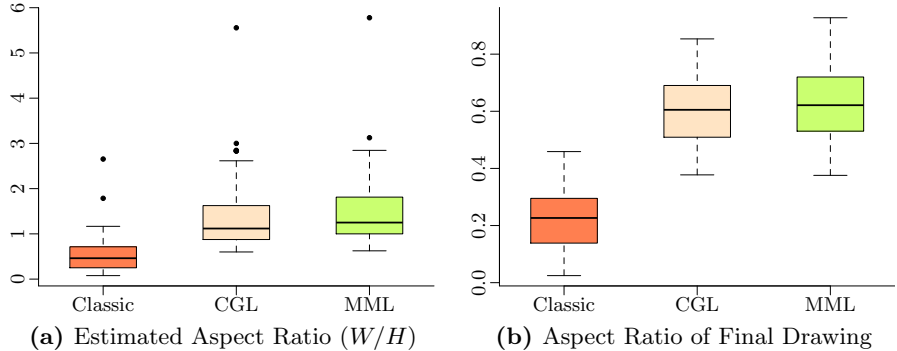
\begin{figure}[tb]
  \centering
  \subfloat[Estimated Aspect Ratio ($W/H$)]{
    \label{fig:results_summary_aspect_ratio_estimated}
    \centering
    \resizebox{.48\textwidth}{!}{%
    \centering
\begin{tikzpicture}[x=1pt,y=1pt]
\definecolor{fillColor}{RGB}{255,255,255}
\path[use as bounding box,fill=fillColor,fill opacity=0.00] (0,0) rectangle (289.08,216.81);
\begin{scope}
\path[clip] ( 24.00, 18.00) rectangle (289.08,204.81);
\definecolor{fillColor}{RGB}{255,127,80}

\path[fill=fillColor] ( 44.97, 30.17) --
	( 89.60, 30.17) --
	( 89.60, 44.26) --
	( 44.97, 44.26) --
	cycle;
\definecolor{drawColor}{RGB}{0,0,0}

\path[draw=drawColor,line width= 1.2pt,line join=round] ( 44.97, 36.59) -- ( 89.60, 36.59);

\path[draw=drawColor,line width= 0.4pt,dash pattern=on 4pt off 4pt ,line join=round,line cap=round] ( 67.29, 24.92) -- ( 67.29, 30.17);

\path[draw=drawColor,line width= 0.4pt,dash pattern=on 4pt off 4pt ,line join=round,line cap=round] ( 67.29, 57.98) -- ( 67.29, 44.26);

\path[draw=drawColor,line width= 0.4pt,line join=round,line cap=round] ( 56.13, 24.92) -- ( 78.44, 24.92);

\path[draw=drawColor,line width= 0.4pt,line join=round,line cap=round] ( 56.13, 57.98) -- ( 78.44, 57.98);

\path[draw=drawColor,line width= 0.4pt,line join=round,line cap=round] ( 44.97, 30.17) --
	( 89.60, 30.17) --
	( 89.60, 44.26) --
	( 44.97, 44.26) --
	( 44.97, 30.17);
\definecolor{fillColor}{RGB}{0,0,0}

\path[draw=drawColor,line width= 0.4pt,line join=round,line cap=round,fill=fillColor] ( 67.29, 76.77) circle (  1.50);

\path[draw=drawColor,line width= 0.4pt,line join=round,line cap=round,fill=fillColor] ( 67.29,103.11) circle (  1.50);
\definecolor{fillColor}{RGB}{255,228,196}

\path[fill=fillColor] (134.23, 49.13) --
	(178.85, 49.13) --
	(178.85, 71.89) --
	(134.23, 71.89) --
	cycle;

\path[draw=drawColor,line width= 1.2pt,line join=round] (134.23, 56.51) -- (178.85, 56.51);

\path[draw=drawColor,line width= 0.4pt,dash pattern=on 4pt off 4pt ,line join=round,line cap=round] (156.54, 40.79) -- (156.54, 49.13);

\path[draw=drawColor,line width= 0.4pt,dash pattern=on 4pt off 4pt ,line join=round,line cap=round] (156.54,101.94) -- (156.54, 71.89);

\path[draw=drawColor,line width= 0.4pt,line join=round,line cap=round] (145.38, 40.79) -- (167.70, 40.79);

\path[draw=drawColor,line width= 0.4pt,line join=round,line cap=round] (145.38,101.94) -- (167.70,101.94);

\path[draw=drawColor,line width= 0.4pt,line join=round,line cap=round] (134.23, 49.13) --
	(178.85, 49.13) --
	(178.85, 71.89) --
	(134.23, 71.89) --
	(134.23, 49.13);
\definecolor{fillColor}{RGB}{0,0,0}

\path[draw=drawColor,line width= 0.4pt,line join=round,line cap=round,fill=fillColor] (156.54,113.61) circle (  1.50);

\path[draw=drawColor,line width= 0.4pt,line join=round,line cap=round,fill=fillColor] (156.54,191.15) circle (  1.50);

\path[draw=drawColor,line width= 0.4pt,line join=round,line cap=round,fill=fillColor] (156.54,108.94) circle (  1.50);

\path[draw=drawColor,line width= 0.4pt,line join=round,line cap=round,fill=fillColor] (156.54,108.55) circle (  1.50);

\path[draw=drawColor,line width= 0.4pt,line join=round,line cap=round,fill=fillColor] (156.54,108.55) circle (  1.50);

\path[draw=drawColor,line width= 0.4pt,line join=round,line cap=round,fill=fillColor] (156.54,108.55) circle (  1.50);
\definecolor{fillColor}{RGB}{202,255,112}

\path[fill=fillColor] (223.48, 52.93) --
	(268.11, 52.93) --
	(268.11, 77.58) --
	(223.48, 77.58) --
	cycle;

\path[draw=drawColor,line width= 1.2pt,line join=round] (223.48, 60.51) -- (268.11, 60.51);

\path[draw=drawColor,line width= 0.4pt,dash pattern=on 4pt off 4pt ,line join=round,line cap=round] (245.79, 41.55) -- (245.79, 52.93);

\path[draw=drawColor,line width= 0.4pt,dash pattern=on 4pt off 4pt ,line join=round,line cap=round] (245.79,108.94) -- (245.79, 77.58);

\path[draw=drawColor,line width= 0.4pt,line join=round,line cap=round] (234.64, 41.55) -- (256.95, 41.55);

\path[draw=drawColor,line width= 0.4pt,line join=round,line cap=round] (234.64,108.94) -- (256.95,108.94);

\path[draw=drawColor,line width= 0.4pt,line join=round,line cap=round] (223.48, 52.93) --
	(268.11, 52.93) --
	(268.11, 77.58) --
	(223.48, 77.58) --
	(223.48, 52.93);
\definecolor{fillColor}{RGB}{0,0,0}

\path[draw=drawColor,line width= 0.4pt,line join=round,line cap=round,fill=fillColor] (245.79,117.40) circle (  1.50);

\path[draw=drawColor,line width= 0.4pt,line join=round,line cap=round,fill=fillColor] (245.79,197.89) circle (  1.50);
\end{scope}
\begin{scope}
\path[clip] (  0.00,  0.00) rectangle (289.08,216.81);
\definecolor{drawColor}{RGB}{0,0,0}

\path[draw=drawColor,line width= 0.4pt,line join=round,line cap=round] ( 67.29, 18.00) -- (245.79, 18.00);

\path[draw=drawColor,line width= 0.4pt,line join=round,line cap=round] ( 67.29, 18.00) -- ( 67.29, 12.00);

\path[draw=drawColor,line width= 0.4pt,line join=round,line cap=round] (156.54, 18.00) -- (156.54, 12.00);

\path[draw=drawColor,line width= 0.4pt,line join=round,line cap=round] (245.79, 18.00) -- (245.79, 12.00);

\node[text=drawColor,anchor=base,inner sep=0pt, outer sep=0pt, scale=  1.50] at ( 67.29,  1.20) {Classic};

\node[text=drawColor,anchor=base,inner sep=0pt, outer sep=0pt, scale=  1.50] at (156.54,  1.20) {CGL};

\node[text=drawColor,anchor=base,inner sep=0pt, outer sep=0pt, scale=  1.50] at (245.79,  1.20) {MML};

\path[draw=drawColor,line width= 0.4pt,line join=round,line cap=round] ( 24.00, 18.00) -- (289.08, 18.00);

\path[draw=drawColor,line width= 0.4pt,line join=round,line cap=round] ( 24.00, 22.58) -- ( 24.00,204.63);

\path[draw=drawColor,line width= 0.4pt,line join=round,line cap=round] ( 24.00, 22.58) -- ( 18.00, 22.58);

\path[draw=drawColor,line width= 0.4pt,line join=round,line cap=round] ( 24.00, 52.93) -- ( 18.00, 52.93);

\path[draw=drawColor,line width= 0.4pt,line join=round,line cap=round] ( 24.00, 83.27) -- ( 18.00, 83.27);

\path[draw=drawColor,line width= 0.4pt,line join=round,line cap=round] ( 24.00,113.61) -- ( 18.00,113.61);

\path[draw=drawColor,line width= 0.4pt,line join=round,line cap=round] ( 24.00,143.95) -- ( 18.00,143.95);

\path[draw=drawColor,line width= 0.4pt,line join=round,line cap=round] ( 24.00,174.29) -- ( 18.00,174.29);

\path[draw=drawColor,line width= 0.4pt,line join=round,line cap=round] ( 24.00,204.63) -- ( 18.00,204.63);

\node[text=drawColor,rotate= 90.00,anchor=base,inner sep=0pt, outer sep=0pt, scale=  1.50] at ( 14.40, 22.58) {0};

\node[text=drawColor,rotate= 90.00,anchor=base,inner sep=0pt, outer sep=0pt, scale=  1.50] at ( 14.40, 52.93) {1};

\node[text=drawColor,rotate= 90.00,anchor=base,inner sep=0pt, outer sep=0pt, scale=  1.50] at ( 14.40, 83.27) {2};

\node[text=drawColor,rotate= 90.00,anchor=base,inner sep=0pt, outer sep=0pt, scale=  1.50] at ( 14.40,113.61) {3};

\node[text=drawColor,rotate= 90.00,anchor=base,inner sep=0pt, outer sep=0pt, scale=  1.50] at ( 14.40,143.95) {4};

\node[text=drawColor,rotate= 90.00,anchor=base,inner sep=0pt, outer sep=0pt, scale=  1.50] at ( 14.40,174.29) {5};

\node[text=drawColor,rotate= 90.00,anchor=base,inner sep=0pt, outer sep=0pt, scale=  1.50] at ( 14.40,204.63) {6};

\path[draw=drawColor,line width= 0.4pt,line join=round,line cap=round] ( 24.00, 18.00) -- ( 24.00,204.63);
\end{scope}
\end{tikzpicture}
    }
  }\hfill
  \subfloat[Aspect Ratio of Final Drawing]{
    \label{fig:results_summary_aspect_ratio_effective}
    \centering
    \resizebox{.48\textwidth}{!}{%
    \centering
\begin{tikzpicture}[x=1pt,y=1pt]
\definecolor{fillColor}{RGB}{255,255,255}
\path[use as bounding box,fill=fillColor,fill opacity=0.00] (0,0) rectangle (289.08,216.81);
\begin{scope}
\path[clip] ( 24.00, 18.00) rectangle (289.08,204.81);
\definecolor{fillColor}{RGB}{255,127,80}

\path[fill=fillColor] ( 44.97, 46.72) --
	( 89.60, 46.72) --
	( 89.60, 76.75) --
	( 44.97, 76.75) --
	cycle;
\definecolor{drawColor}{RGB}{0,0,0}

\path[draw=drawColor,line width= 1.2pt,line join=round] ( 44.97, 63.59) -- ( 89.60, 63.59);

\path[draw=drawColor,line width= 0.4pt,dash pattern=on 4pt off 4pt ,line join=round,line cap=round] ( 67.29, 24.92) -- ( 67.29, 46.72);

\path[draw=drawColor,line width= 0.4pt,dash pattern=on 4pt off 4pt ,line join=round,line cap=round] ( 67.29,108.11) -- ( 67.29, 76.75);

\path[draw=drawColor,line width= 0.4pt,line join=round,line cap=round] ( 56.13, 24.92) -- ( 78.44, 24.92);

\path[draw=drawColor,line width= 0.4pt,line join=round,line cap=round] ( 56.13,108.11) -- ( 78.44,108.11);

\path[draw=drawColor,line width= 0.4pt,line join=round,line cap=round] ( 44.97, 46.72) --
	( 89.60, 46.72) --
	( 89.60, 76.75) --
	( 44.97, 76.75) --
	( 44.97, 46.72);
\definecolor{fillColor}{RGB}{255,228,196}

\path[fill=fillColor] (134.23,117.75) --
	(178.85,117.75) --
	(178.85,152.46) --
	(134.23,152.46) --
	cycle;

\path[draw=drawColor,line width= 1.2pt,line join=round] (134.23,136.14) -- (178.85,136.14);

\path[draw=drawColor,line width= 0.4pt,dash pattern=on 4pt off 4pt ,line join=round,line cap=round] (156.54, 92.50) -- (156.54,117.75);

\path[draw=drawColor,line width= 0.4pt,dash pattern=on 4pt off 4pt ,line join=round,line cap=round] (156.54,183.71) -- (156.54,152.46);

\path[draw=drawColor,line width= 0.4pt,line join=round,line cap=round] (145.38, 92.50) -- (167.70, 92.50);

\path[draw=drawColor,line width= 0.4pt,line join=round,line cap=round] (145.38,183.71) -- (167.70,183.71);

\path[draw=drawColor,line width= 0.4pt,line join=round,line cap=round] (134.23,117.75) --
	(178.85,117.75) --
	(178.85,152.46) --
	(134.23,152.46) --
	(134.23,117.75);
\definecolor{fillColor}{RGB}{202,255,112}

\path[fill=fillColor] (223.48,121.82) --
	(268.11,121.82) --
	(268.11,158.15) --
	(223.48,158.15) --
	cycle;

\path[draw=drawColor,line width= 1.2pt,line join=round] (223.48,139.23) -- (268.11,139.23);

\path[draw=drawColor,line width= 0.4pt,dash pattern=on 4pt off 4pt ,line join=round,line cap=round] (245.79, 92.17) -- (245.79,121.82);

\path[draw=drawColor,line width= 0.4pt,dash pattern=on 4pt off 4pt ,line join=round,line cap=round] (245.79,197.89) -- (245.79,158.15);

\path[draw=drawColor,line width= 0.4pt,line join=round,line cap=round] (234.64, 92.17) -- (256.95, 92.17);

\path[draw=drawColor,line width= 0.4pt,line join=round,line cap=round] (234.64,197.89) -- (256.95,197.89);

\path[draw=drawColor,line width= 0.4pt,line join=round,line cap=round] (223.48,121.82) --
	(268.11,121.82) --
	(268.11,158.15) --
	(223.48,158.15) --
	(223.48,121.82);
\end{scope}
\begin{scope}
\path[clip] (  0.00,  0.00) rectangle (289.08,216.81);
\definecolor{drawColor}{RGB}{0,0,0}

\path[draw=drawColor,line width= 0.4pt,line join=round,line cap=round] ( 67.29, 18.00) -- (245.79, 18.00);

\path[draw=drawColor,line width= 0.4pt,line join=round,line cap=round] ( 67.29, 18.00) -- ( 67.29, 12.00);

\path[draw=drawColor,line width= 0.4pt,line join=round,line cap=round] (156.54, 18.00) -- (156.54, 12.00);

\path[draw=drawColor,line width= 0.4pt,line join=round,line cap=round] (245.79, 18.00) -- (245.79, 12.00);

\node[text=drawColor,anchor=base,inner sep=0pt, outer sep=0pt, scale=  1.50] at ( 67.29,  1.20) {Classic};

\node[text=drawColor,anchor=base,inner sep=0pt, outer sep=0pt, scale=  1.50] at (156.54,  1.20) {CGL};

\node[text=drawColor,anchor=base,inner sep=0pt, outer sep=0pt, scale=  1.50] at (245.79,  1.20) {MML};

\path[draw=drawColor,line width= 0.4pt,line join=round,line cap=round] ( 24.00, 18.00) -- (289.08, 18.00);

\path[draw=drawColor,line width= 0.4pt,line join=round,line cap=round] ( 24.00, 20.19) -- ( 24.00,173.48);

\path[draw=drawColor,line width= 0.4pt,line join=round,line cap=round] ( 24.00, 20.19) -- ( 18.00, 20.19);

\path[draw=drawColor,line width= 0.4pt,line join=round,line cap=round] ( 24.00, 58.51) -- ( 18.00, 58.51);

\path[draw=drawColor,line width= 0.4pt,line join=round,line cap=round] ( 24.00, 96.83) -- ( 18.00, 96.83);

\path[draw=drawColor,line width= 0.4pt,line join=round,line cap=round] ( 24.00,135.16) -- ( 18.00,135.16);

\path[draw=drawColor,line width= 0.4pt,line join=round,line cap=round] ( 24.00,173.48) -- ( 18.00,173.48);

\node[text=drawColor,rotate= 90.00,anchor=base,inner sep=0pt, outer sep=0pt, scale=  1.50] at ( 14.40, 20.19) {0.0};

\node[text=drawColor,rotate= 90.00,anchor=base,inner sep=0pt, outer sep=0pt, scale=  1.50] at ( 14.40, 58.51) {0.2};

\node[text=drawColor,rotate= 90.00,anchor=base,inner sep=0pt, outer sep=0pt, scale=  1.50] at ( 14.40, 96.83) {0.4};

\node[text=drawColor,rotate= 90.00,anchor=base,inner sep=0pt, outer sep=0pt, scale=  1.50] at ( 14.40,135.16) {0.6};

\node[text=drawColor,rotate= 90.00,anchor=base,inner sep=0pt, outer sep=0pt, scale=  1.50] at ( 14.40,173.48) {0.8};

\path[draw=drawColor,line width= 0.4pt,line join=round,line cap=round] ( 24.00, 18.00) -- ( 24.00,204.81);
\end{scope}
\end{tikzpicture}
    }
  }
  \caption{
    Comparison of the aspect ratio of the \attar instances
    when laid out with traditional methods~\cite{EadesLS93,GansnerKNV93}
    (box on the left), CGL (middle), and MML (right).
    It can be seen that the methods presented here
    clearly improve the aspect ratio of the final~drawing.
  }
  \label{fig:results_summary_aspect_ratio}
\end{figure}
\paragraph{(H3): Aspect Ratio.}
Exemplary drawings of both \acs{cgl} and \acs{mml} can be seen in
\autoref{fig:mainexample} and \autoref{fig:appendixexample}.
Both show layouts of \attar instances.
Whereas the aspect ratio of the original layout of \autoref{fig:mainexample}
is about $0.14$, the ratio of
the new layouts is about $0.6$.
This improvement of the aspect ratio has been
achieved by reversing three arcs (now pointing upwards). The created drawings
are significantly more compact. In \autoref{fig:mainexample_3}, generated with
the MML model, the reversed arcs can be found easily, since the model tries
to maximize their length.

An average of about $3.13$ arcs needed to be reversed on the \attar graphs
and $3.82$ on the random graphs to adhere to the selected $H$.
Also the maximum number of reversed arcs in both benchmark sets is
similar; it is $8$ for the \attar instances and $9$ for the random instances.
The reversed arcs changed the estimated aspect ratio
of the \attar graphs from an average of $0.51$
to an average of $1.36$, see \autoref{fig:results_summary_aspect_ratio_estimated}.
As mentioned earlier, the estimated aspect ratio must not necessarily coincide with the
final drawing's aspect ratio.
To further inspect this, we produced final drawings
using the same strategies of the Sugiyama approach as discussed in~\cite{RueeggESvH16}.
In \autoref{fig:results_summary_aspect_ratio_effective},
one can see that the average aspect ratio improves from $0.22$
to about $0.61$ for CGL and to about $0.63$ for MML.
From this we conclude that both models lead to compact layouts with improved
aspect~ratio.

\section{Conclusion}
\label{sec:conclusion}

This paper introduces the \acs{cglp}, which can be seen as an extension of
the DAG Layering Problem suggested by Healy and Nikolov \cite{HealyN02b,HealyN02a}, and
the \acs{glp} suggested by R{\"u}egg \etal~\cite{RueeggESvH16}.
The \acs{cglp} gives more control over the desired layering
by integrating the reversal of arcs and taking the contribution
of dummy vertices to a layering's width into account.

We suggest two MIP models for \acs{cglp},
one of which is based on partial orderings
and show that the model can be solved
to optimality within a short computation time
for typical instances with up to $100$ vertices.
In addition, we suggest an alternative MIP model (MML)
for a slightly different problem which can be solved even faster while
the widths and arc lengths of the generated layerings
do not increase significantly.
Our experiments have shown that using the CGLP, indeed, the
aspect ratio of the generated final drawings can be influenced.

\subsubsection{\textbf{Acknowledgements.}}
This work was supported by the German Research Foundation under
the project \emph{Compact Graph Drawing with Port Constraints}
(ComDraPor, DFG HA 4407/8-1 and MU 1129/9-1).

\bibliographystyle{abbrv}
\bibliography{cau-rt,pub-rts,newrefs}

\begin{thebibliography}{10}

\bibitem{CoffmanG72}
E.~G. Coffman. and R.~L. Graham.
\newblock Optimal scheduling for two-processor systems.
\newblock {\em Acta Informatica}, 1(3):200--213, 1972.

\bibitem{EadesLS93}
P.~Eades, X.~Lin, and W.~F. Smyth.
\newblock A fast and effective heuristic for the feedback arc set problem.
\newblock {\em Information Processing Letters}, 47(6):319--323, 1993.

\bibitem{EadesS90}
P.~Eades and K.~Sugiyama.
\newblock How to draw a directed graph.
\newblock {\em Journal of Information Processing}, 13(4):424--437, 1990.

\bibitem{GansnerKNV93}
E.~R. Gansner, E.~Koutsofios, S.~C. North, and K.-P. Vo.
\newblock A technique for drawing directed graphs.
\newblock {\em Software Engineering}, 19(3):214--230, 1993.

\bibitem{HealyN02b}
P.~Healy and N.~S. Nikolov.
\newblock A branch-and-cut approach to the directed acyclic graph layering
  problem.
\newblock In {\em Proceedings of the 10th International Symposium on Graph
  Drawing (GD'02)}, volume 2528 of {\em LNCS}, pages 98--109. Springer, 2002.

\bibitem{HealyN02a}
P.~Healy and N.~S. Nikolov.
\newblock How to layer a directed acyclic graph.
\newblock In {\em Proceedings of the 9th International Symposium on Graph
  Drawing (GD'01)}, volume 2265 of {\em LNCS}, pages 563--566. Springer, 2002.

\bibitem{NachmansonRL08}
L.~Nachmanson, G.~Robertson, and B.~Lee.
\newblock Drawing graphs with {GLEE}.
\newblock In S.-H. Hong, T.~Nishizeki, and W.~Quan, editors, {\em Graph
  Drawing}, volume 4875 of {\em LNCS}, pages 389--394. Springer Berlin
  Heidelberg, 2008.

\bibitem{NikolovTB05}
N.~S. Nikolov, A.~Tarassov, and J.~Branke.
\newblock In search for efficient heuristics for minimum-width graph layering
  with consideration of dummy nodes.
\newblock {\em Journal of Experimental Algorithmics}, 10, 2005.

\bibitem{RueeggESvH16}
U.~R{\"u}egg, T.~Ehlers, M.~Sp{\"o}nemann, and R.~von Hanxleden.
\newblock A generalization of the directed graph layering problem.
\newblock In {\em Proceedings of the 24th International Symposium on Graph
  Drawing and Network Visualization (GD'16)}, 2016.

\bibitem{SugiyamaTT81}
K.~Sugiyama, S.~Tagawa, and M.~Toda.
\newblock Methods for visual understanding of hierarchical system structures.
\newblock {\em {IEEE} Transactions on Systems, Man and Cybernetics},
  11(2):109--125, Feb. 1981.

\bibitem{Wilf67}
H.~S. Wilf.
\newblock The eigenvalues of a graph and its chromatic number.
\newblock {\em Journal London Math.\ Soc.}, 42:330--332, 1967.

\end{thebibliography}

\newpage
\appendix
\section{Appendix}

\subsection{Additional Experimental Results}

\subsubsection{Example Layout of an AT\&T Instance}
\begin{figure}[h!]
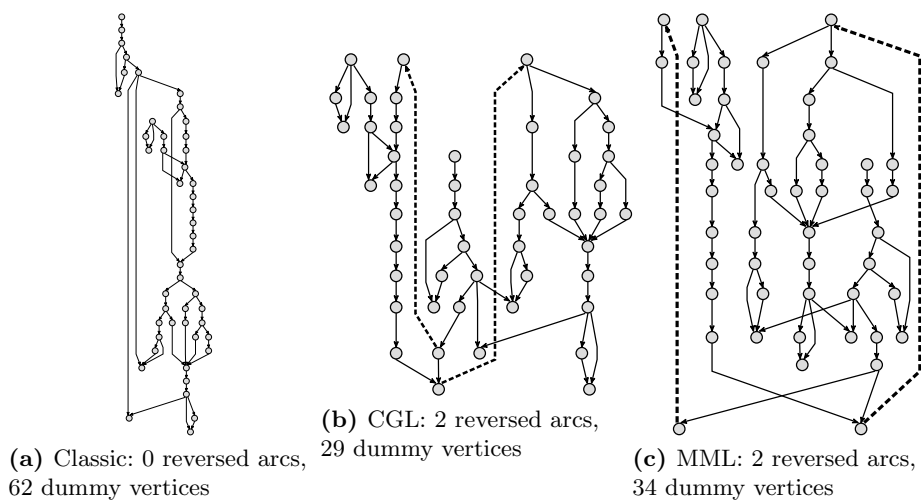

  \centering
  \begin{minipage}[b]{0.32\textwidth}
    \centering
    \subfloat[][Classic: 0 reversed arcs,\\62 dummy vertices]{
      \label{fig:appendixexample_1}
      \begin{minipage}{\textwidth}
        \centering
        \includegraphics[scale=0.165]{{{data/layerings/drawings/g.48.6_d62_classic}}}
      \end{minipage}
    }
  \end{minipage}
  \hfill
  \begin{minipage}[b]{0.32\textwidth}
    \centering
    \subfloat[][CGL: 2 reversed arcs,\\29 dummy vertices]{
    \label{fig:appendixexample_2}
      \begin{minipage}{\textwidth}
        \centering
        \includegraphics[scale=0.32]{{{data/layerings/drawings/g.48.6_G}}}
      \end{minipage}
    }
  \end{minipage}
  \hfill
  \begin{minipage}[b]{0.32\textwidth}
    \centering
    \subfloat[][MML: 2 reversed arcs,\\34 dummy vertices]{
	  \label{fig:appendixexample_3}
      \begin{minipage}{\textwidth}
        \centering
        \includegraphics[scale=0.32]{{{data/layerings/drawings/g.48.6_H}}}
      \end{minipage}
    }
  \end{minipage}
  \caption{
    (a) Layout with existing methods where every arc has to point downwards.
     (b) Layout with solution of model CGL
   (c) Layout with solution of model MML. The reversed arcs in (b) and (c) are drawn bold and
   dashed.
  }
  \label{fig:appendixexample}
\end{figure}

\autoref{fig:appendixexample} shows an instance of the \mbox{\attar} benchmark set
with 48 vertices drawn with the standard layering method \cite{EadesLS93,GansnerKNV93}.
If it is required that all arcs point downward, then the layering will have a poor aspect ratio
as shown in \autoref{fig:appendixexample_1}.
If we allow to reverse only two arcs, the aspect ratio can be improved tremendously
(see \autoref{fig:appendixexample_2} and \ref{fig:appendixexample_3}).
The methods CGL and MML, presented in this paper,
are able to respect a bound on the height of the drawing,
allowing a number of arcs to point upwards.
The created drawings are also significantly more compact.
The reversed arcs in (b) and (c) are drawn bold and dashed.

\subsubsection{More Results of the CGL and MML Models}

\begin{figure}[tb]
  \centering
  \subfloat[Execution Time]{
    \includegraphics[width=.48\textwidth]{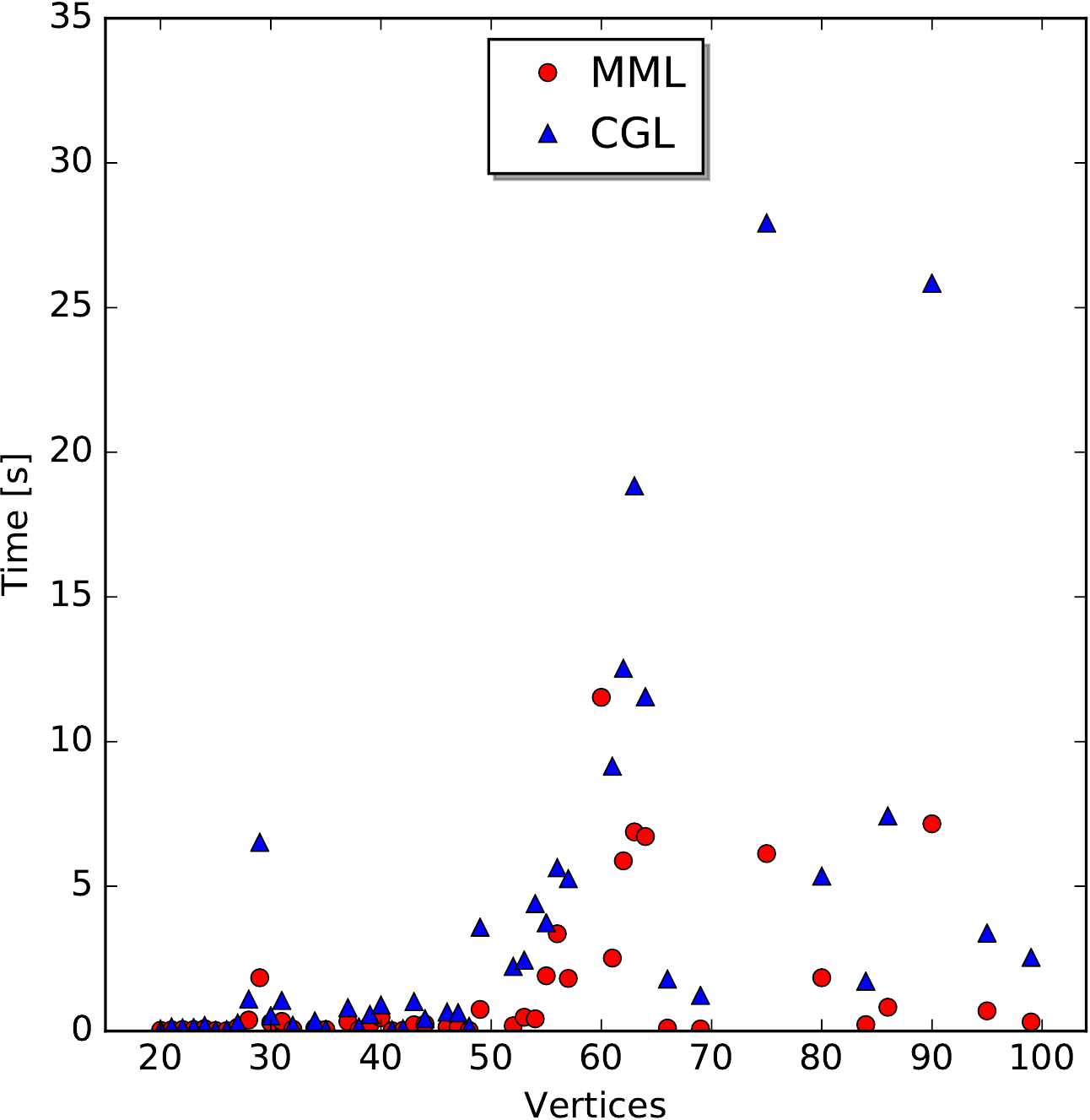}
  }\\
  \subfloat[Arc Length]{
    \includegraphics[width=.48\textwidth]{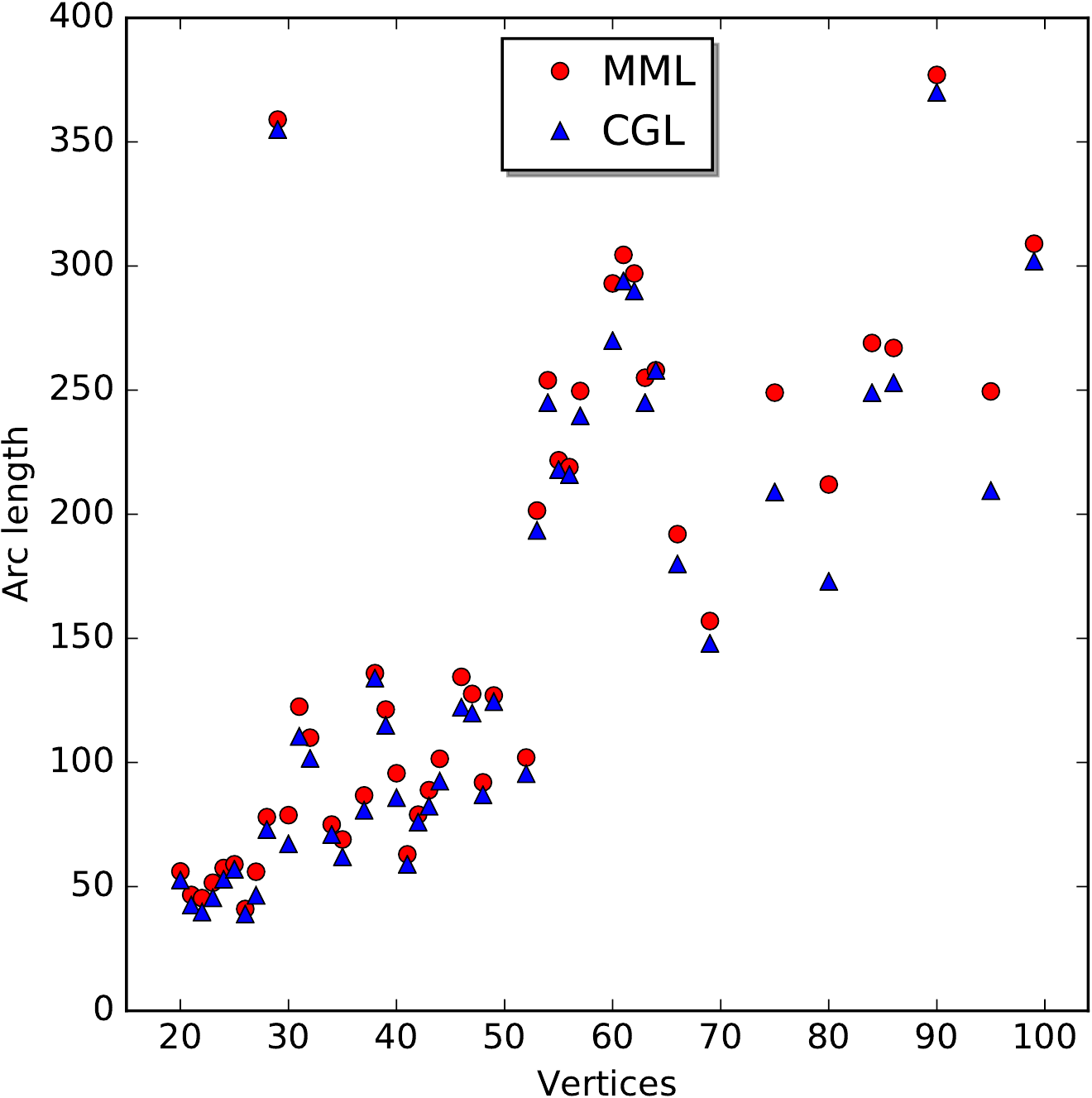}
  }\hfill
  \subfloat[W]{
    \includegraphics[width=.48\textwidth]{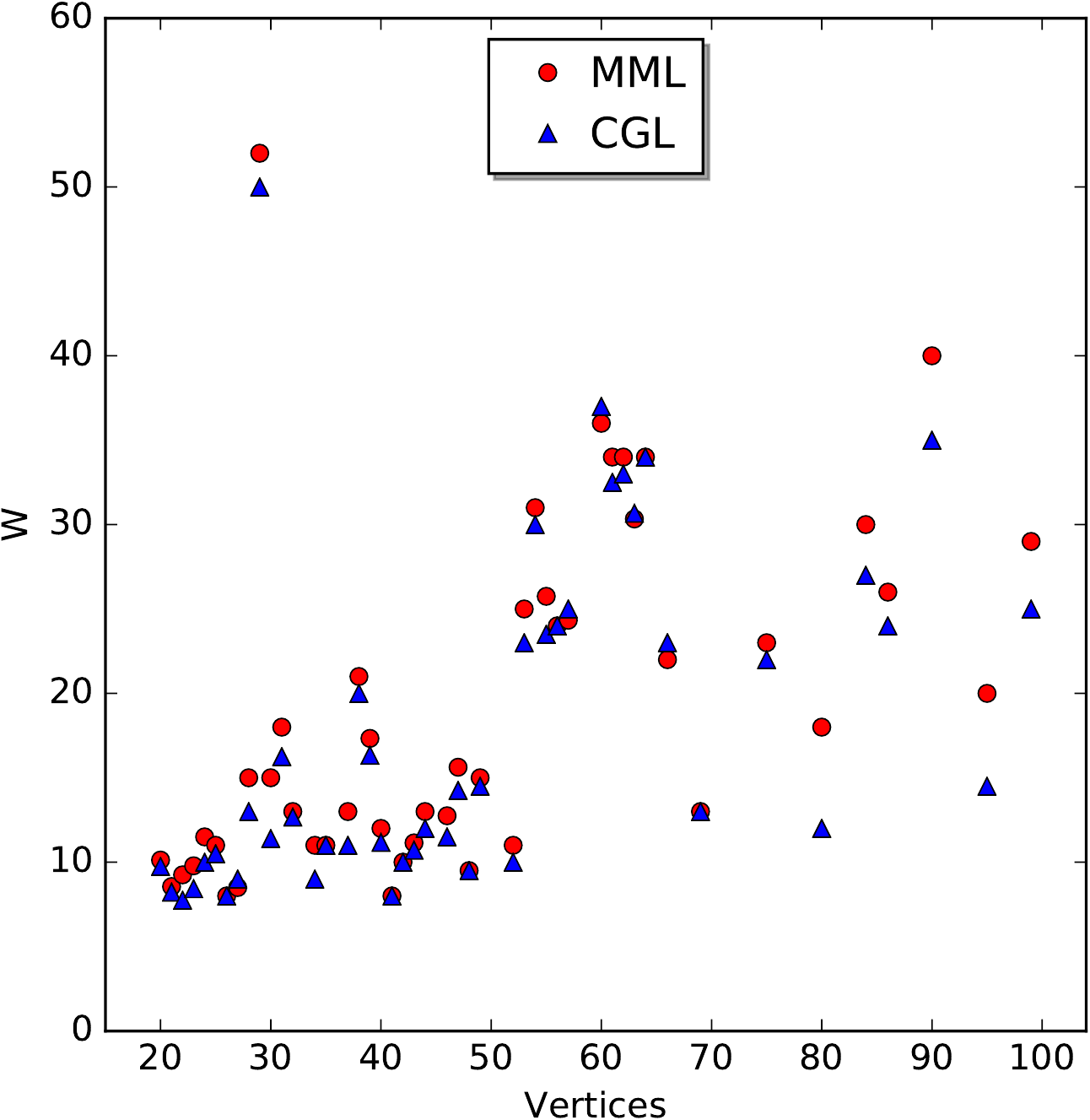}
  }
  \caption{Results for the models CGL and MML for the \attar graphs. Values for instances with the same number of vertices are averaged.}
  \label{fig:results_scatter_att}
\end{figure}

\begin{figure}[tb]
  \centering
  \subfloat[Execution Time]{
    \includegraphics[width=.48\textwidth]{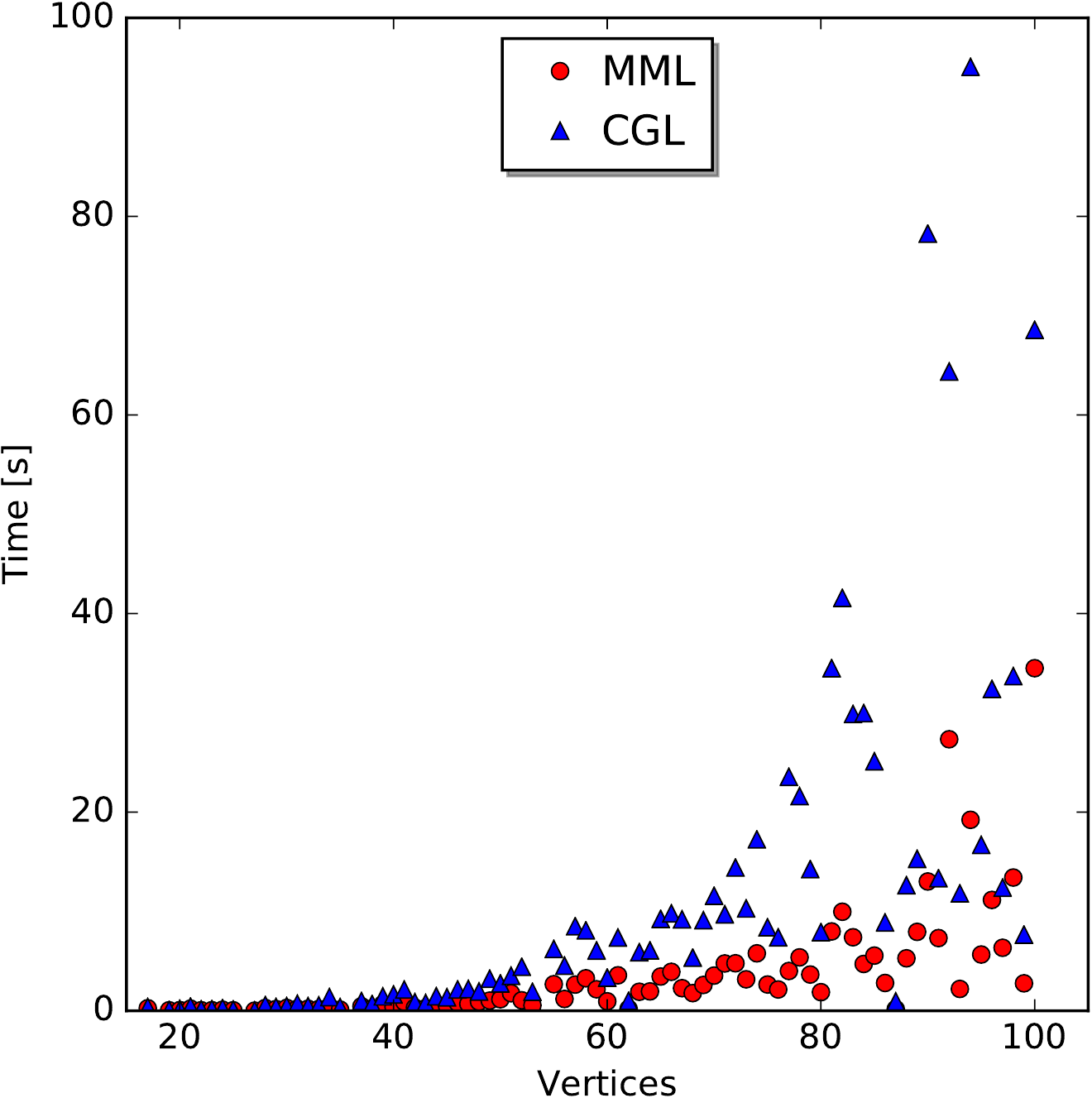}
  }\\
  \subfloat[Arc Length]{
    \includegraphics[width=.48\textwidth]{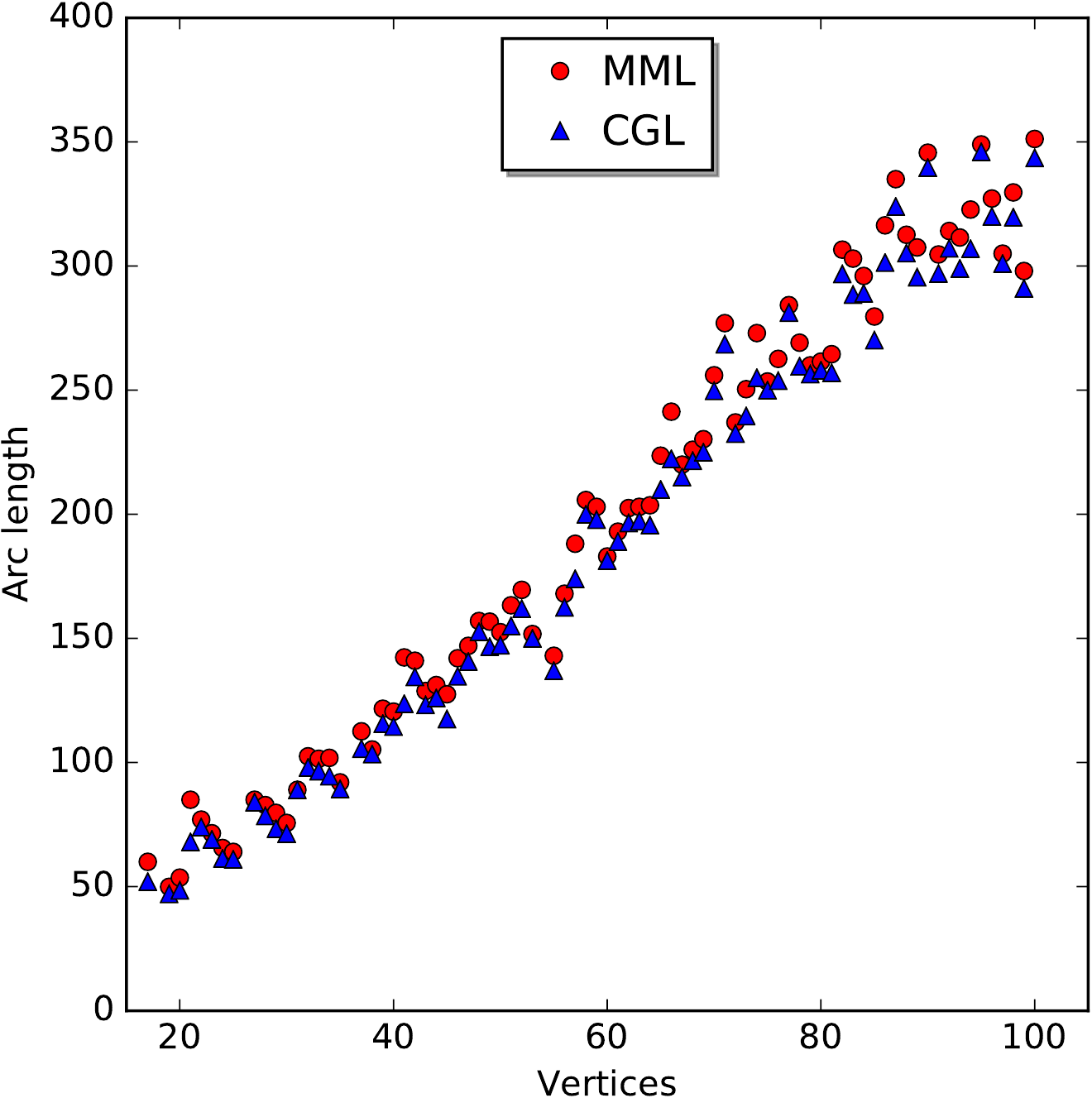}
  }\hfill
  \subfloat[W]{
    \includegraphics[width=.48\textwidth]{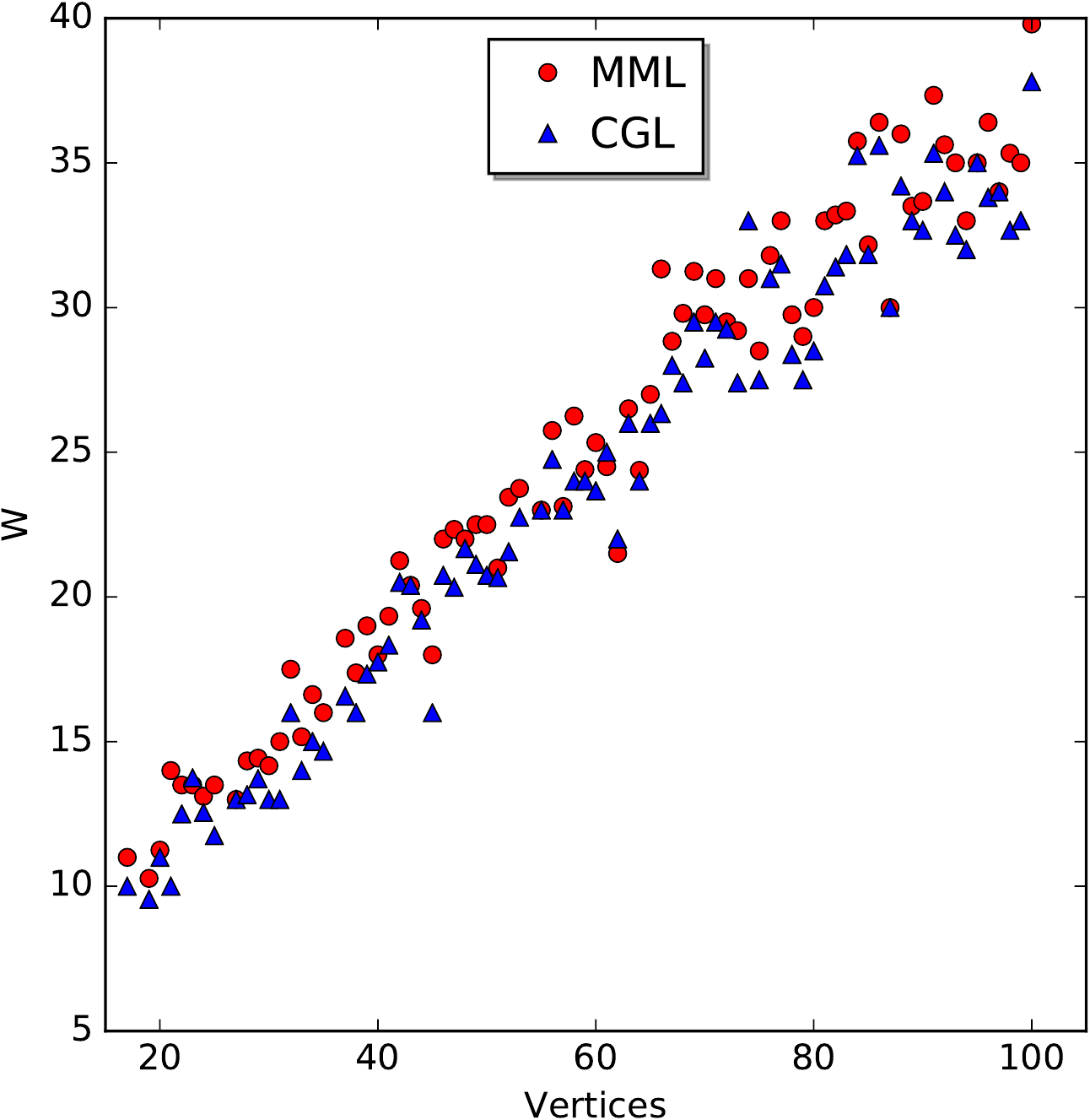}
  }
  \caption{
     Results for the models CGL and MML for the random graphs. Values for instances with the same number of vertices are averaged.}
  \label{fig:results_scatter_random}
\end{figure}

Tables \ref{tab:summary_table_att} and \ref{tab:summary_table_rand} show the results for the benchmark sets \attar and Random. In both tables, $\mathcal{I}$ denotes the number of instances for the given vertex count range. $\bar{n}$ and $\bar{e}$ are the average number of vertices and arcs. $\overline{W}$ is the average width, \ie, the maximum number of vertices per layer including dummy vertices. For each drawing, the height was fixed as discussed in the paper. $\overline{EL}$ and $\overline{RE}$ denote the average arc lengths and number of reversed arcs. Standard deviations are given in brackets.
A good overview over the distribution of the runtimes across the instances can also be obtained from
\autoref{fig:results_scatter_att} and \autoref{fig:results_scatter_random}.

\setlength{\tabcolsep}{.5em}
%
\begin{table}[tb]
\centering
\caption{Summary table for the set of \attar graphs. Notations are defined above.} 
\label{tab:summary_table_att}
\begin{tabular}{lrrrlrrr}
  \toprule
Node Counts & $\mathcal{I}$ & $\bar{n}$ & $\bar{e}$ & $$ & $\overline{W}$ & $\overline{EL}$ & $\overline{RE}$ \\ 
  \midrule
\multirow{2}{*}{[15, 30)}\enspace & \multirow{2}{*}{53} & \multirow{2}{*}{22.7 {\scriptsize[2.1]}} & \multirow{2}{*}{33.3 {\scriptsize[16.4]}} & CGL & 9.6 {\scriptsize[6.5]} & 52.6 {\scriptsize[46.8]} & 2.1 {\scriptsize[1.3]} \\ 
   &  &  &  & MML & 10.6 {\scriptsize[6.8]} & 57.5 {\scriptsize[46.7]} & 2.1 {\scriptsize[1.3]} \\ 
   \midrule\multirow{2}{*}{[30, 45)}\enspace & \multirow{2}{*}{44} & \multirow{2}{*}{38.0 {\scriptsize[5.1]}} & \multirow{2}{*}{52.8 {\scriptsize[10.8]}} & CGL & 12.2 {\scriptsize[4.3]} & 88.1 {\scriptsize[30.8]} & 2.9 {\scriptsize[1.3]} \\ 
   &  &  &  & MML & 13.4 {\scriptsize[4.6]} & 96.2 {\scriptsize[29.6]} & 2.9 {\scriptsize[1.3]} \\ 
   \midrule\multirow{2}{*}{[45, 60)}\enspace & \multirow{2}{*}{29} & \multirow{2}{*}{50.5 {\scriptsize[4.1]}} & \multirow{2}{*}{80.8 {\scriptsize[26.9]}} & CGL & 17.1 {\scriptsize[7.7]} & 155.2 {\scriptsize[68.5]} & 3.8 {\scriptsize[1.5]} \\ 
   &  &  &  & MML & 18.2 {\scriptsize[7.4]} & 162.5 {\scriptsize[68.8]} & 3.8 {\scriptsize[1.5]} \\ 
   \midrule\multirow{2}{*}{[60, 75)}\enspace & \multirow{2}{*}{12} & \multirow{2}{*}{62.8 {\scriptsize[2.6]}} & \multirow{2}{*}{112.6 {\scriptsize[12.0]}} & CGL & 30.2 {\scriptsize[6.8]} & 254.8 {\scriptsize[49.1]} & 5.7 {\scriptsize[1.2]} \\ 
   &  &  &  & MML & 30.5 {\scriptsize[7.2]} & 265.0 {\scriptsize[48.4]} & 5.7 {\scriptsize[1.2]} \\ 
   \midrule\multirow{2}{*}{[75, 90)}\enspace & \multirow{2}{*}{4} & \multirow{2}{*}{81.2 {\scriptsize[4.9]}} & \multirow{2}{*}{117.0 {\scriptsize[15.8]}} & CGL & 21.2 {\scriptsize[6.5]} & 221.0 {\scriptsize[37.7]} & 5.2 {\scriptsize[1.7]} \\ 
   &  &  &  & MML & 24.2 {\scriptsize[5.1]} & 249.2 {\scriptsize[26.4]} & 5.2 {\scriptsize[1.7]} \\ 
   \midrule\multirow{2}{*}{[90, 105)}\enspace & \multirow{2}{*}{4} & \multirow{2}{*}{94.8 {\scriptsize[3.7]}} & \multirow{2}{*}{139.5 {\scriptsize[15.1]}} & CGL & 22.2 {\scriptsize[9.8]} & 272.8 {\scriptsize[78.3]} & 5.5 {\scriptsize[0.6]} \\ 
   &  &  &  & MML & 27.2 {\scriptsize[9.5]} & 296.2 {\scriptsize[60.8]} & 5.5 {\scriptsize[0.6]} \\ 
   \bottomrule
\end{tabular}
\end{table}

\begin{table}[tb]
\centering
\caption{Summary table for the set of random graphs. Notations are defined above.} 
\label{tab:summary_table_rand}
\begin{tabular}{lrrrlrrr}
  \toprule
Node Counts & $\mathcal{I}$ & $\bar{n}$ & $\bar{e}$ & $$ & $\overline{W}$ & $\overline{EL}$ & $\overline{RE}$ \\ 
  \midrule
\multirow{2}{*}{[15, 30)}\enspace & \multirow{2}{*}{54} & \multirow{2}{*}{23.3 {\scriptsize[3.7]}} & \multirow{2}{*}{36.9 {\scriptsize[6.0]}} & CGL & 11.9 {\scriptsize[2.1]} & 61.4 {\scriptsize[12.7]} & 2.5 {\scriptsize[1.1]} \\ 
   &  &  &  & MML & 12.6 {\scriptsize[2.1]} & 65.7 {\scriptsize[13.3]} & 2.5 {\scriptsize[1.1]} \\ 
   \midrule\multirow{2}{*}{[30, 45)}\enspace & \multirow{2}{*}{65} & \multirow{2}{*}{37.2 {\scriptsize[4.3]}} & \multirow{2}{*}{59.0 {\scriptsize[7.6]}} & CGL & 16.5 {\scriptsize[3.1]} & 105.3 {\scriptsize[19.7]} & 3.0 {\scriptsize[1.4]} \\ 
   &  &  &  & MML & 17.7 {\scriptsize[2.8]} & 111.0 {\scriptsize[22.1]} & 3.0 {\scriptsize[1.4]} \\ 
   \midrule\multirow{2}{*}{[45, 60)}\enspace & \multirow{2}{*}{62} & \multirow{2}{*}{52.4 {\scriptsize[4.3]}} & \multirow{2}{*}{82.9 {\scriptsize[7.1]}} & CGL & 22.0 {\scriptsize[3.1]} & 159.8 {\scriptsize[27.1]} & 4.0 {\scriptsize[1.6]} \\ 
   &  &  &  & MML & 23.1 {\scriptsize[3.0]} & 167.4 {\scriptsize[29.0]} & 4.0 {\scriptsize[1.6]} \\ 
   \midrule\multirow{2}{*}{[60, 75)}\enspace & \multirow{2}{*}{60} & \multirow{2}{*}{66.7 {\scriptsize[3.9]}} & \multirow{2}{*}{105.9 {\scriptsize[7.0]}} & CGL & 26.7 {\scriptsize[3.6]} & 216.3 {\scriptsize[31.8]} & 4.4 {\scriptsize[1.5]} \\ 
   &  &  &  & MML & 27.8 {\scriptsize[3.8]} & 223.8 {\scriptsize[32.4]} & 4.4 {\scriptsize[1.5]} \\ 
   \midrule\multirow{2}{*}{[75, 90)}\enspace & \multirow{2}{*}{63} & \multirow{2}{*}{81.6 {\scriptsize[4.2]}} & \multirow{2}{*}{128.7 {\scriptsize[7.5]}} & CGL & 31.4 {\scriptsize[4.2]} & 277.3 {\scriptsize[31.7]} & 4.4 {\scriptsize[1.6]} \\ 
   &  &  &  & MML & 32.6 {\scriptsize[4.0]} & 286.1 {\scriptsize[32.8]} & 4.4 {\scriptsize[1.6]} \\ 
   \midrule\multirow{2}{*}{[90, 105)}\enspace & \multirow{2}{*}{36} & \multirow{2}{*}{94.6 {\scriptsize[3.3]}} & \multirow{2}{*}{147.9 {\scriptsize[5.1]}} & CGL & 34.1 {\scriptsize[3.4]} & 316.9 {\scriptsize[38.5]} & 5.2 {\scriptsize[1.7]} \\ 
   &  &  &  & MML & 35.9 {\scriptsize[3.2]} & 325.3 {\scriptsize[36.3]} & 5.2 {\scriptsize[1.7]} \\ 
   \bottomrule
\end{tabular}
\end{table}

\end{document}